\documentclass[prl,twocolumn,amssymb,amsmath,nobibnotes,aps,showpacs]{revtex4-1}

\usepackage{graphics}
\usepackage{graphicx}
\usepackage{braket}
\usepackage{float}
\usepackage{multirow}

\usepackage[usenames,dvipsnames]{color}

\begin{document}

\title{Feshbach spectroscopy of an ultracold mixture of $^{85}$Rb and $^{133}$Cs}

\author{Hung-Wen Cho$^{1}$, Daniel J. McCarron$^{1}$, Michael P. K\"oppinger$^{1}$,
Daniel L. Jenkin$^{1}$, Kirsteen L. Butler$^{1}$, Paul~S.~Julienne$^{2}$,
Caroline L. Blackley$^{3}$, C. Ruth Le Sueur$^{3}$, Jeremy M. Hutson$^{3}$, and
Simon L. Cornish$^{1}$}

\affiliation{$^{1}$Joint Quantum Centre (JQC) Durham/Newcastle, Department of
Physics, Durham University, South Road, Durham DH1 3LE, United Kingdom}
\affiliation{$^{2}$Joint Quantum Institute, NIST and the University of
Maryland, Gaithersburg, Maryland 20899-8423, USA} \affiliation{$^{3}$Joint
Quantum Centre (JQC) Durham/Newcastle, Department of Chemistry, Durham
University, Durham DH1 3LE, United Kingdom}

\date{\today}

\begin{abstract}
We report the observation of interspecies Feshbach resonances in an optically
trapped mixture of $^{85}$Rb and $^{133}$Cs. We measure 14 interspecies
features in the lowest spin channels for a magnetic field range from 0 to 700~G
and show that they are in good agreement with coupled-channel calculations. The
interspecies background scattering length is close to zero over a large range
of magnetic fields, permitting the sensitive detection of Feshbach resonances
through interspecies thermalisation. Our results confirm the quality of the
Rb-Cs potential curves and offer promising starting points for the production
of ultracold polar molecules.
\end{abstract}


\maketitle

Ultracold polar molecules provide numerous new and exciting avenues of research
for studies of dilute quantum systems \cite{Carr:NJPintro:2009,Friedrich2009}.
The permanent electric dipole moments possessed by these molecules give rise to
anisotropic, long-range dipole-dipole interactions which are in contrast to the
isotropic, short-range interactions commonly encountered in ultracold atomic
gas experiments \cite{Lahaye2009}. These dipole-dipole interactions can operate
over a range greater than typical optical lattice separations leading to a
range of novel quantum phases and opportunities for quantum simulation and
quantum information processing
\cite{Capogrosso-Sansone2010,Micheli:2007,Wall2009}.

The most promising route towards realizing these proposals exploits a two-step
indirect method where the constituent atoms in a mixed-species quantum gas are
associated into ground-state molecules \cite{Damski:2003}. Weakly bound
molecules are first made by magneto-association using a Feshbach resonance
\cite{Chin:RMP:2010} and are then optically transferred into the rovibrational
ground state by stimulated Raman adiabatic passage (STIRAP)
\cite{Bergmann1998}. Great strides have been made using this approach in a
number of systems \cite{Ni:KRb:2008,Lang:ground:2008,Danzl:ground:2010},
although the only polar molecule that has so far been produced at high
phase-space density is fermionic KRb \cite{Ni:KRb:2008}. However, two KRb
molecules can undergo an exothermic reaction to form K$_2$ + Rb$_2$
\cite{Ospelkaus:react:2010}. An attractive alternative is to form ground-state
RbCs, which is expected to be collisionally stable because both the exchange
reaction \mbox{2RbCs $\rightarrow$ Rb$_2$ + Cs$_2$} and trimer formation
reactions are endothermic \cite{Zuchowski2010}. There has been considerable
work on the Feshbach resonances \cite{Pilch:2009} and molecule formation
\cite{Debatin:2011, Takekoshi:RbCs:2012} in $^{87}$Rb$^{133}$Cs. However, this
isotopologue has an interspecies background scattering length that is large and
positive, which produces a spatial separation of the dual condensate
\cite{McCarron2011} and enhances losses from 3-body collisions \cite{Cho2011,
Lercher:2011}. Both of these factors inhibit the formation of weakly bound
Feshbach molecules.

In this Letter we explore the alternative mixture of $^{85}$Rb and
$^{133}$Cs, which we show does not suffer from the problems present for
the mixture of $^{87}$Rb and $^{133}$Cs. We report the observation of
14 interspecies Feshbach resonances in excellent agreement with
coupled-channel calculations. We show that the interspecies background
scattering length is close to zero over a large range of magnetic
fields, permitting the sensitive detection of Feshbach resonances
through interspecies thermalisation. Our observations together with
detailed calculations of the near-threshold bound-state spectrum reveal
numerous possible gateways into the realm of ultracold heteronuclear
molecules.

Details of our apparatus have been described previously in the context of our
work on dual-species condensates of  $^{87}$Rb and $^{133}$Cs
\cite{McCarron2011,Cho2011}. Ultracold mixtures of $^{85}$Rb and $^{133}$Cs are
collected in a two-species magneto-optical trap. The $^{85}$Rb and $^{133}$Cs
atoms are optically pumped into the $|2,-2\rangle$ and $|3,-3\rangle$ states,
respectively, and then loaded into a magnetic quadrupole trap. Forced RF
evaporation cools the $^{85}$Rb atoms to $50~\mu$K while $^{133}$Cs is cooled
sympathetically by interspecies elastic collisions. Further efficient
evaporative cooling is inhibited by Majorana losses \cite{Lin2009}. The two
species are then transferred into a crossed dipole trap formed using the output
of a 30\,W, 1550\,nm fibre laser. After loading, the $^{85}$Rb and $^{133}$Cs
atoms are transferred into the $|2,+2\rangle$ and $|3,+3\rangle$ states
respectively by RF adiabatic rapid passage \cite{Jenkin2011} and a vertical
magnetic field gradient of 21.2~G/cm is applied, just below the 22.4~G/cm required to levitate $^{85}$Rb
\footnote{Units of gauss rather than tesla, the accepted SI unit for the
magnetic field, have been used in this paper to conform to the conventional
usage in this field of physics.}. The resulting gravitational sag of the
$^{133}$Cs cloud ($\lesssim2$~$\mu$m) is significantly less than the typical
vertical FWHM of the cloud ($\simeq24\ \mu$m), so that there is excellent
spatial overlap of both species throughout the measurement. The magnetic field
gradient also results in a field spread of $\simeq0.05~$G across the cloud,
which limits the minimum observed resonance width. The magnetic field is
calibrated using microwave spectroscopy between the hyperfine states of
$^{133}$Cs.

\begin{figure*}[htb]
\centering
\includegraphics[width=2\columnwidth]{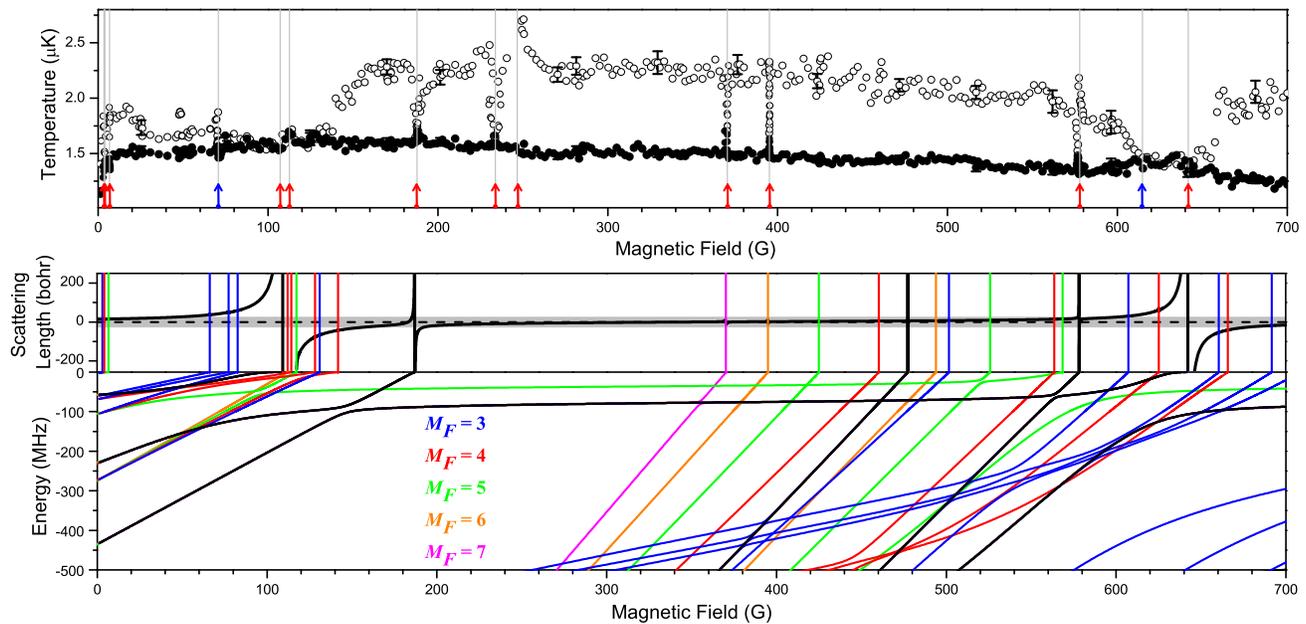}
\caption{(Color online) Top: Observation of interspecies Feshbach
resonances by thermalisation of the two species for fields up to 700~G.
The closed (open) symbols indicate the $^{85}$Rb ($^{133}$Cs)
temperature while red (blue) arrows mark observed $s$-wave ($p$-wave)
resonances. Error bars show the standard deviations for multiple
control shots at specific magnetic fields. Bottom: The calculated
$s$-wave interspecies scattering length and weakly bound molecular
states for the same field range. Bound states arising from $L=0$ ($s$
states) are shown as black lines while those from $L=2$ ($d$ states)
are shown in other colors that indicate the value of $M_{F}$ (see
figure). The bound-state energies are plotted relative to the energy of
the lowest hyperfine state of $^{85}$Rb + $^{133}$Cs, the (2,2) + (3,3)
hyperfine level. Resonance positions are marked on the scattering
length plot using vertical lines of the same color as the corresponding
bound state. The grey shaded region indicates where $|a|<30$~bohr. All
the bound states shown are for $M_{\rm tot}=5$, corresponding to s-wave
scattering in the lowest channel.} \label{fig:BroadScan}
\end{figure*}

A typical experiment starts with a mixture of \mbox{$2.0(1)\times10^{5}$}
$^{85}$Rb atoms at $7.9(1)~\mu$K and $2.3(1)\times10^{4}$ $^{133}$Cs
atoms at $10.6(5)~\mu$K confined in the dipole trap in the lowest spin
channels. The temperature difference between the two species arises
from a combination of the small interspecies background scattering
length and the differing trap depths for the two species: at 1550~nm
the polarisability of $^{133}$Cs is $\sim1.4$ times greater than that
of $^{85}$Rb. The significant atom number imbalance between the two
species increases the sensitivity of heteronuclear Feshbach
spectroscopy. Here the $^{133}$Cs atoms act as a probe species immersed
in a collisional bath of $^{85}$Rb \cite{Wille2008}. To perform
Feshbach spectroscopy, the magnetic field is switched to a specific
value in the range 0 to 700~G. Evaporative cooling is then performed by
reducing the laser powers by a factor of 4 over 2~s to final trap
depths of 15~$\mu$K for $^{85}$Rb and 22~$\mu$K for $^{133}$Cs. The
mixture is then held for 1~s in this final potential. To probe the
mixture, resonant absorption images of both species are captured in
each experimental cycle using a frame-transfer CCD camera. Interspecies
Feshbach resonances are identified by studying the variation in the
atom number and temperature for both species with magnetic field.

The rich Feshbach resonance structure of this system is shown in Fig.\
\ref{fig:BroadScan}. The top panel shows a coarse scan of the $^{85}$Rb
($\bullet$) and $^{133}$Cs ($\circ$) temperatures between 0 and 700~G.
Each point corresponds to an average of at least 3 repeated
measurements. The bottom panel shows the scattering length and the
energies of the near-threshold molecular states, obtained from
coupled-channel calculations as described below. Over a large range of
magnetic field the magnitude of the interspecies scattering length is
$<30~$bohr (indicated by the grey shaded region) and the two species do
not equilibrate to a common temperature within the duration of the
experiment. However, the variation of the scattering length in the
vicinity of an interspecies resonance produces a pronounced feature in
the $^{133}$Cs temperature as the $^{133}$Cs atoms are sympathetically
cooled by the colder $^{85}$Rb atoms. The large windows of thermal
equilibrium around 110~G and 640~G coincide with broad $s$-wave
resonances. Additionally, many narrow resonances can be identified in
very good agreement with the calculated Feshbach spectrum. The
positions of experimental resonances are marked by arrows whose
positions are determined by fine scans across each resonance. Red
arrows mark $s$-wave features while blue arrows mark two observed
$p$-wave resonances.

\begin{figure}[b]
\centering
\includegraphics[width=1\columnwidth]{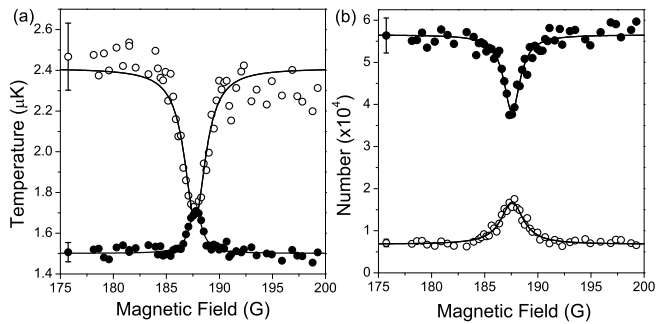}
\caption{An interspecies resonance observed at 187.66(5)~G. The closed
(open) symbols indicate $^{85}$Rb ($^{133}$Cs) data points. (a)
Temperature data for both species reveals the thermalisation of
$^{133}$Cs with $^{85}$Rb at and near the resonance highlighting the
tunable nature of the sympathetic cooling. (b) Atom number data show
the $^{133}$Cs number to increase while the $^{85}$Rb number decreases
when a resonance is crossed and sympathetic cooling occurs. Error bars
show the standard deviation for multiple control shots at a specific
magnetic field.} \label{fig:187G}
\end{figure}

Fig.\ \ref{fig:187G} shows an example of a fine scan across the
interspecies resonance at 187.66(5)~G. Here the thermalisation between
the two species on resonance is clear as $^{133}$Cs is sympathetically
cooled by $^{85}$Rb (Fig.\ \ref{fig:187G}(a)). The sympathetic cooling
also enhances the $^{133}$Cs number remaining in the trap while the
additional heat load on $^{85}$Rb leads to further evaporative trap
loss (Fig.\ \ref{fig:187G}(b)). The lines correspond to Lorentzian fits
to the data. In general, the experimental positions ($B_{0}$) and
widths ($\delta$) of the resonances are determined from weighted
averages of the fits to the 4 features illustrated in Fig.\
\ref{fig:187G}. However, in some cases fewer than 4 fits are possible
because only the number or temperature results show pronounced
features: for example, only the number results show pronounced features
in the broad window of thermal equilibrium near 110~G.

\begin{figure}[t]
\centering
\includegraphics[width=1\columnwidth]{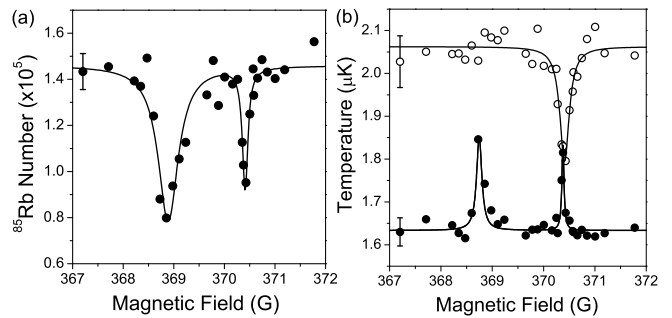}
\caption{An intraspecies $^{85}$Rb resonance at 368.78(3)~G neighbouring an
interspecies resonance at 370.39(1)~G. Closed (open) symbols indicate $^{85}$Rb
($^{133}$Cs) data. (a) Loss of $^{85}$Rb reveals both resonances. (b)
Temperature data allows the resonances to be distinguished as intra-or
interspecies. The $^{133}$Cs temperature is unchanged at the $^{85}$Rb alone
resonance in contrast to the interspecies resonance at higher field due to
interspecies collisions causing sympathetic cooling. Error bars show the
standard deviation for multiple control shots at a specific magnetic field.}
\label{fig:370G}
\end{figure}

Intra- and interspecies Feshbach resonances are distinguished as shown
in Fig.\ \ref{fig:370G}. In this region, the $^{85}$Rb number shows two
loss features (Fig.\ \ref{fig:370G}(a)). The resonance at 368.78(3)~G
results in no change of the $^{133}$Cs temperature, indicating that
this feature is a $^{85}$Rb intraspecies resonance. Conversely, near
the feature at 370.39(1)~G, the two species come into thermal
equilibrium through sympathetic cooling, showing this to be an
interspecies resonance (Fig.\ \ref{fig:370G}(b)). This interpretation
is confirmed by performing single-species measurements. Further details
on the observed $^{85}$Rb intraspecies resonances will be presented in
ref.\ \cite{Blackley2012}.

The interspecies scattering length and bound-state positions are
calculated from a coupled-channel model, using the potential curves of
ref.\ \cite{Takekoshi:RbCs:2012}, which were fitted to Fourier
transform spectra of both $^{85}$Rb$^{133}$Cs and $^{87}$Rb$^{133}$Cs
and Feshbach resonances and weakly bound states of $^{87}$Rb$^{133}$Cs.
All the calculations are carried out in a fully uncoupled basis set,
$\ket{s_{\rm{Rb}}m_{s\rm{Rb}}}\ket{i_{\rm{Rb}}m_{i\rm{Rb}}}
\ket{s_{\rm{Cs}}m_{s\rm{Cs}}}\ket{i_{\rm{Cs}}m_{i\rm{Cs}}}\ket{LM_{L}}$,
where $s$ and $i$ indicate electron and nuclear spins and $L$ is the
quantum number for end-over-end rotation of the two atoms about one
another. The coupled equations are diagonal in the total projection
quantum number $M_{\rm tot}= M_{L}+M_{\rm{F}}$, where $M_{\rm{F}} =
m_{s\rm{Rb}}+m_{i\rm{Rb}}+m_{s\rm{Cs}}+m_{i\rm{Cs}}$. The basis sets
used here included all functions with $L=0$ and 2 for incoming $L=0$
($s$-wave) and with $L=1$ and 3 for incoming $L=1$ ($p$-wave).

The scattering calculations are carried out using the MOLSCAT program
\cite{molscat:v14-short}, as modified to handle collisions in an
external field \cite{Gonzalez-Martinez:2007}. The calculations use a
fixed-step log-derivative propagator \cite{Manolopoulos:1986} to
propagate from 0.3 to 1.9~nm and then the variable-step Airy propagator
\cite{Alexander:1984} to propagate from 1.9 to 1,500~nm. The $s$-wave
scattering length is obtained from $a(k) = (ik)^{-1}
(1-S_{00})/(1+S_{00})$ \cite{Hutson:res:2007}, where $S_{00}$ is the
diagonal S-matrix element in the incoming channel and $k$ is the
corresponding wavevector. For incoming $L=1$, $a$ and $k$ are replaced
by $a^3$ and $k^3$. Single-channel calculations on the singlet and
triplet potential curves of ref.\ \cite{Takekoshi:RbCs:2012} give
singlet and triplet scattering lengths of 585.6 and 11.27 bohr
respectively.

The bound-state calculations use the associated packages BOUND
\cite{Hutson:bound-short:1993} and FIELD, which locate bound states by solving
sets of coupled differential equations in the same basis set as for scattering
calculations, as described for alkali-metal dimers in ref.\
\cite{Hutson:Cs2:2008}. BOUND locates the energies of bound states at fixed
magnetic field, whereas FIELD locates the magnetic fields at which bound states
exist at fixed binding energy.

In the vicinity of a resonance, the scattering length is given approximately by
$a(B)=a_{\rm bg}\left(1-(\Delta/(B-B_0))\right)$. The width $\Delta$ is thus
conveniently calculated for each resonance as $B_*-B_0$, where $B_*$ is the
field where $a(B)$ crosses zero and $B_0$ is the location of the corresponding
pole. The MOLSCAT package is capable of converging directly on both these
points. However, it should be noted that theoretical widths defined in this way
are not the same thing as the fitted experimental Lorentzian widths $\delta$,
and the two should not be compared.

\begin{table}[htb]
\begin{center}
\begin{tabular}{ccccccccccc}
\hline
\hline
 \multicolumn{2}{c} {Experiment}& & \multicolumn{8}{c} {Theory}\\
 \cline{1-2}
 \cline{4-11}
 $B_0$ & $\delta$ & & \multicolumn{4}{c} {Assignment} & $B_0$ & $B_*$ & $\Delta$ & $a_{\rm{bg}}$\\
 \cline{4-7}
  (G) & (G) & & $L_{\rm i}$ & $L$ & $F$ & $M_{F}$ & (G) & (G) & (G) &  (bohr) \\
 \hline
 3.47(1) & 0.12(2) & & $s$ & 2 & 5& 3& 3.10 & 3.10& 0.00005 & 27.5 \\   
 3.90(2) & 0.33(5) & & $s$ & 2& 5& 4& 4.27 & 4.27&0.00029 & 27.8 \\
 6.76(2) & 0.14(3) & & $s$ & 2& 5& 5& 6.80 & 6.80&0.00086 & 28.6 \\
 70.68(4)& 0.8(1) & & $p$ & 1& -&- & 70.54 & 58.54 & -12& -\\
	& & & $s$ & 2 & 4& 3& 77.51 & 77.52 & 0.010 & 93.6 \\
 107.13(1) & 0.6(2) & & $s$ & 0& 5& 5& 109 & 350 &  241 & 9.6 \\
 112.6(4) & 28(5) & & $s$ &  2& 6& 6& 112.29 &112.12 &  -0.17& -628 \\
	& & & $s$ & 2 & 4& 4& 114.33 &114.21 & -0.12 & -246 \\
	& & & $s$ & 2 & 6& 5& 117.40 &117.35 & -0.051 & -169 \\
 187.66(5) & 1.7(3) & & $s$ & 0& 6&5 & 187.07 & 182.97& -4.1 & -30.3 \\
 233.9(2) & 2.1(3) & &- & -&- &- &- &- \\
 246.5(3) & 14(2) & & -&- &- & -& -& -\\
 370.39(1) & 0.08(4) & & $s$ & 2&7 & 7& 370.41 &374.31 & 3.9 & 1.57 \\
 395.20(1) & 0.08(1) & & $s$ & 2 &7 & 6& 395.11 & 395.56& 0.45 & 3.4 \\
 	& & & $s$ & 2 & 7& 5&425.11 &425.16 & 0.045 & 6.1  \\
	& & & $s$ & 2 & 5& 5&568.62 &568.66 & 0.037 & 29.8 \\
 577.8(1) & 1.1(3) & & $s$ & 0&6& 5& 578.36 &578.7 0& 0.34 & 32.2 \\
 614.6(3) & 1.1(4) & & $p$ & 1& -&- & 614.98 & 608.18 & -6.8 & -\\
	& & & $s$ & 2 & 5& 4&625.29 &625.30& 0.014 & 123 \\
 641.8(3) & 6(2) & & $s$ &  0 &5 & 5& 642 & 901 &  259 & 9.6  \\
	& & & $s$ & 2 & 3 & 3&708.70 &708.68& -0.024 & -23.8  \\
\hline
\hline
\end{tabular}
\end{center}
\caption[]{Feshbach resonances for $^{85}$Rb $|2,+2\rangle$ + $^{133}$Cs
$|3,+3\rangle$ in the field range 0 to 700~G. All resonances with calculated
widths $\Delta>0.01~$G are listed. See Supplemental Material at {\em [URL will
be inserted by publisher]} for a complete listing of the $s$-wave resonances,
including narrower ones. The experimental errors $\delta$ are statistical
uncertainties resulting from the fits as described in the text. Additional
systematic uncertainties of $0.1$~G and $0.5$~G apply to the experimental
resonance positions in the field ranges 0 to 400~G and 400 to 700~G
respectively.} \label{table:85RbCsFeshbachResonances}
\end{table}

The experimental and theoretical resonance parameters are compared in Table
\ref{table:85RbCsFeshbachResonances}. Most of the experimentally observed
resonance positions are in good agreement with the theoretical predictions
\footnote{Comparison between theory and experiment for the cluster of
resonances near 110~G is limited by the fact that not all resonances are
resolved experimentally. Two experimental observations, strong thermalisation
at 233.9(2)~G and strong loss at 246.5(3)~G, do not correspond to 2-body
resonances using the current theoretical model and require further
investigation.}. The success of mass scaling between $^{87}$Rb$^{133}$Cs and
$^{85}$Rb$^{133}$Cs demonstrates the accuracy of the potential curves, and in
particular confirms that they support the correct absolute number of bound
states. This agreement extends over features measured for both incoming $s$-
and $p$-wave collisions \footnote{The two $p$-wave resonances that are observed
experimentally have calculated widths $>6$~G, and all others have calculated
widths $<0.3$~G.}.

Our observations show that the interspecies background scattering length is
close to zero over a large range of magnetic fields. This reduces losses due to
interspecies 3-body collisions for $^{85}$Rb + $^{133}$Cs and makes it much
easier to achieve good overlap of atomic clouds than for $^{87}$Rb +
$^{133}$Cs. The broad resonances near 110 G and 640 G will permit precise
tuning of the interatomic interactions, allowing control of the miscibility and
studies of Efimov physics in heteronuclear systems. There are also numerous
narrower resonances with widths in the range that is convenient for
magnetoassociation. Future work will include exploration of the molecular bound
states (Fig.\ \ref{fig:BroadScan}) by magnetic-field modulation spectroscopy,
magnetoassociation to form weakly bound molecules, and optical transfer to the
ground state. $^{85}$Rb$^{133}$Cs thus offers the prospect of forming a gas of
ultracold polar molecules that are stable with respect to chemical reactions.

This work was supported by the UK EPSRC, by AFOSR MURI Grant FA9550-09-1-0617,
and by EOARD Grant FA8655-10-1-3033.

\bibliography{Rb85Cs,../all}

\begin{thebibliography}{34}%
\makeatletter
\providecommand \@ifxundefined [1]{%
 \@ifx{#1\undefined}
}%
\providecommand \@ifnum [1]{%
 \ifnum #1\expandafter \@firstoftwo
 \else \expandafter \@secondoftwo
 \fi
}%
\providecommand \@ifx [1]{%
 \ifx #1\expandafter \@firstoftwo
 \else \expandafter \@secondoftwo
 \fi
}%
\providecommand \natexlab [1]{#1}%
\providecommand \enquote  [1]{``#1''}%
\providecommand \bibnamefont  [1]{#1}%
\providecommand \bibfnamefont [1]{#1}%
\providecommand \citenamefont [1]{#1}%
\providecommand \href@noop [0]{\@secondoftwo}%
\providecommand \href [0]{\begingroup \@sanitize@url \@href}%
\providecommand \@href[1]{\@@startlink{#1}\@@href}%
\providecommand \@@href[1]{\endgroup#1\@@endlink}%
\providecommand \@sanitize@url [0]{\catcode `\\12\catcode `\$12\catcode
  `\&12\catcode `\#12\catcode `\^12\catcode `\_12\catcode `\%12\relax}%
\providecommand \@@startlink[1]{}%
\providecommand \@@endlink[0]{}%
\providecommand \url  [0]{\begingroup\@sanitize@url \@url }%
\providecommand \@url [1]{\endgroup\@href {#1}{\urlprefix }}%
\providecommand \urlprefix  [0]{URL }%
\providecommand \Eprint [0]{\href }%
\providecommand \doibase [0]{http://dx.doi.org/}%
\providecommand \selectlanguage [0]{\@gobble}%
\providecommand \bibinfo  [0]{\@secondoftwo}%
\providecommand \bibfield  [0]{\@secondoftwo}%
\providecommand \translation [1]{[#1]}%
\providecommand \BibitemOpen [0]{}%
\providecommand \bibitemStop [0]{}%
\providecommand \bibitemNoStop [0]{.\EOS\space}%
\providecommand \EOS [0]{\spacefactor3000\relax}%
\providecommand \BibitemShut  [1]{\csname bibitem#1\endcsname}%
\let\auto@bib@innerbib\@empty
\bibitem [{\citenamefont {Carr}\ \emph {et~al.}(2009)\citenamefont {Carr},
  \citenamefont {{DeMille}}, \citenamefont {Krems},\ and\ \citenamefont
  {Ye}}]{Carr:NJPintro:2009}%
  \BibitemOpen
  \bibfield  {author} {\bibinfo {author} {\bibfnamefont {L.~D.}\ \bibnamefont
  {Carr}}, \bibinfo {author} {\bibfnamefont {D.}~\bibnamefont {{DeMille}}},
  \bibinfo {author} {\bibfnamefont {R.~V.}\ \bibnamefont {Krems}}, \ and\
  \bibinfo {author} {\bibfnamefont {J.}~\bibnamefont {Ye}},\ }\href@noop {}
  {\bibfield  {journal} {\bibinfo  {journal} {New J. Phys.}\ }\textbf {\bibinfo
  {volume} {11}},\ \bibinfo {pages} {055049} (\bibinfo {year}
  {2009})}\BibitemShut {NoStop}%
\bibitem [{\citenamefont {Friedrich}\ and\ \citenamefont
  {Doyle}(2009)}]{Friedrich2009}%
  \BibitemOpen
  \bibfield  {author} {\bibinfo {author} {\bibfnamefont {B.}~\bibnamefont
  {Friedrich}}\ and\ \bibinfo {author} {\bibfnamefont {J.~M.}\ \bibnamefont
  {Doyle}},\ }\href@noop {} {\bibfield  {journal} {\bibinfo  {journal}
  {ChemPhysChem}\ }\textbf {\bibinfo {volume} {10}},\ \bibinfo {pages} {604}
  (\bibinfo {year} {2009})}\BibitemShut {NoStop}%
\bibitem [{\citenamefont {Lahaye}\ \emph {et~al.}(2009)\citenamefont {Lahaye},
  \citenamefont {Menotti}, \citenamefont {Santos}, \citenamefont {Lewenstein},\
  and\ \citenamefont {Pfau}}]{Lahaye2009}%
  \BibitemOpen
  \bibfield  {author} {\bibinfo {author} {\bibfnamefont {T.}~\bibnamefont
  {Lahaye}}, \bibinfo {author} {\bibfnamefont {C.}~\bibnamefont {Menotti}},
  \bibinfo {author} {\bibfnamefont {L.}~\bibnamefont {Santos}}, \bibinfo
  {author} {\bibfnamefont {M.}~\bibnamefont {Lewenstein}}, \ and\ \bibinfo
  {author} {\bibfnamefont {T.}~\bibnamefont {Pfau}},\ }\href@noop {} {\bibfield
   {journal} {\bibinfo  {journal} {Rep. Prog. Phys.}\ }\textbf {\bibinfo
  {volume} {72}},\ \bibinfo {pages} {126401} (\bibinfo {year}
  {2009})}\BibitemShut {NoStop}%
\bibitem [{\citenamefont {Capogrosso-Sansone}\ \emph
  {et~al.}(2010)\citenamefont {Capogrosso-Sansone}, \citenamefont {Trefzger},
  \citenamefont {Lewenstein}, \citenamefont {Zoller},\ and\ \citenamefont
  {Pupillo}}]{Capogrosso-Sansone2010}%
  \BibitemOpen
  \bibfield  {author} {\bibinfo {author} {\bibfnamefont {B.}~\bibnamefont
  {Capogrosso-Sansone}}, \bibinfo {author} {\bibfnamefont {C.}~\bibnamefont
  {Trefzger}}, \bibinfo {author} {\bibfnamefont {M.}~\bibnamefont
  {Lewenstein}}, \bibinfo {author} {\bibfnamefont {P.}~\bibnamefont {Zoller}},
  \ and\ \bibinfo {author} {\bibfnamefont {G.}~\bibnamefont {Pupillo}},\
  }\href@noop {} {\bibfield  {journal} {\bibinfo  {journal} {Phys. Rev. Lett.}\
  }\textbf {\bibinfo {volume} {104}},\ \bibinfo {pages} {125301} (\bibinfo
  {year} {2010})}\BibitemShut {NoStop}%
\bibitem [{\citenamefont {Micheli}\ \emph {et~al.}(2007)\citenamefont
  {Micheli}, \citenamefont {Pupillo}, \citenamefont {B\"uchler},\ and\
  \citenamefont {Zoller}}]{Micheli:2007}%
  \BibitemOpen
  \bibfield  {author} {\bibinfo {author} {\bibfnamefont {A.}~\bibnamefont
  {Micheli}}, \bibinfo {author} {\bibfnamefont {G.}~\bibnamefont {Pupillo}},
  \bibinfo {author} {\bibfnamefont {H.~P.}\ \bibnamefont {B\"uchler}}, \ and\
  \bibinfo {author} {\bibfnamefont {P.}~\bibnamefont {Zoller}},\ }\href@noop {}
  {\bibfield  {journal} {\bibinfo  {journal} {Phys. Rev. A}\ }\textbf {\bibinfo
  {volume} {76}},\ \bibinfo {pages} {043604} (\bibinfo {year}
  {2007})}\BibitemShut {NoStop}%
\bibitem [{\citenamefont {Wall}\ and\ \citenamefont {Carr}(2009)}]{Wall2009}%
  \BibitemOpen
  \bibfield  {author} {\bibinfo {author} {\bibfnamefont {M.~L.}\ \bibnamefont
  {Wall}}\ and\ \bibinfo {author} {\bibfnamefont {L.~D.}\ \bibnamefont
  {Carr}},\ }\href@noop {} {\bibfield  {journal} {\bibinfo  {journal} {New J.
  Phys.}\ }\textbf {\bibinfo {volume} {11}},\ \bibinfo {pages} {055027}
  (\bibinfo {year} {2009})}\BibitemShut {NoStop}%
\bibitem [{\citenamefont {Damski}\ \emph {et~al.}(2003)\citenamefont {Damski},
  \citenamefont {Santos}, \citenamefont {Tiemann}, \citenamefont {Lewenstein},
  \citenamefont {Kotochigova}, \citenamefont {Julienne},\ and\ \citenamefont
  {Zoller}}]{Damski:2003}%
  \BibitemOpen
  \bibfield  {author} {\bibinfo {author} {\bibfnamefont {B.}~\bibnamefont
  {Damski}}, \bibinfo {author} {\bibfnamefont {L.}~\bibnamefont {Santos}},
  \bibinfo {author} {\bibfnamefont {E.}~\bibnamefont {Tiemann}}, \bibinfo
  {author} {\bibfnamefont {M.}~\bibnamefont {Lewenstein}}, \bibinfo {author}
  {\bibfnamefont {S.}~\bibnamefont {Kotochigova}}, \bibinfo {author}
  {\bibfnamefont {P.}~\bibnamefont {Julienne}}, \ and\ \bibinfo {author}
  {\bibfnamefont {P.}~\bibnamefont {Zoller}},\ }\href@noop {} {\bibfield
  {journal} {\bibinfo  {journal} {Phys. Rev. Lett.}\ }\textbf {\bibinfo
  {volume} {90}},\ \bibinfo {pages} {110401} (\bibinfo {year}
  {2003})}\BibitemShut {NoStop}%
\bibitem [{\citenamefont {Chin}\ \emph {et~al.}(2010)\citenamefont {Chin},
  \citenamefont {Grimm}, \citenamefont {Tiesinga},\ and\ \citenamefont
  {Julienne}}]{Chin:RMP:2010}%
  \BibitemOpen
  \bibfield  {author} {\bibinfo {author} {\bibfnamefont {C.}~\bibnamefont
  {Chin}}, \bibinfo {author} {\bibfnamefont {R.}~\bibnamefont {Grimm}},
  \bibinfo {author} {\bibfnamefont {E.}~\bibnamefont {Tiesinga}}, \ and\
  \bibinfo {author} {\bibfnamefont {P.~S.}\ \bibnamefont {Julienne}},\
  }\href@noop {} {\bibfield  {journal} {\bibinfo  {journal} {Rev. Mod. Phys.}\
  }\textbf {\bibinfo {volume} {82}},\ \bibinfo {pages} {1225} (\bibinfo {year}
  {2010})}\BibitemShut {NoStop}%
\bibitem [{\citenamefont {Bergmann}\ \emph {et~al.}(1998)\citenamefont
  {Bergmann}, \citenamefont {Theuer},\ and\ \citenamefont
  {Shore}}]{Bergmann1998}%
  \BibitemOpen
  \bibfield  {author} {\bibinfo {author} {\bibfnamefont {K.}~\bibnamefont
  {Bergmann}}, \bibinfo {author} {\bibfnamefont {H.}~\bibnamefont {Theuer}}, \
  and\ \bibinfo {author} {\bibfnamefont {B.~W.}\ \bibnamefont {Shore}},\
  }\href@noop {} {\bibfield  {journal} {\bibinfo  {journal} {Rev. Mod. Phys.}\
  }\textbf {\bibinfo {volume} {70}},\ \bibinfo {pages} {1003} (\bibinfo {year}
  {1998})}\BibitemShut {NoStop}%
\bibitem [{\citenamefont {Ni}\ \emph {et~al.}(2008)\citenamefont {Ni},
  \citenamefont {Ospelkaus}, \citenamefont {{de Miranda}}, \citenamefont
  {Pe'er}, \citenamefont {Neyenhuis}, \citenamefont {Zirbel}, \citenamefont
  {Kotochigova}, \citenamefont {Julienne}, \citenamefont {Jin},\ and\
  \citenamefont {Ye}}]{Ni:KRb:2008}%
  \BibitemOpen
  \bibfield  {author} {\bibinfo {author} {\bibfnamefont {K.-K.}\ \bibnamefont
  {Ni}}, \bibinfo {author} {\bibfnamefont {S.}~\bibnamefont {Ospelkaus}},
  \bibinfo {author} {\bibfnamefont {M.~H.~G.}\ \bibnamefont {{de Miranda}}},
  \bibinfo {author} {\bibfnamefont {A.}~\bibnamefont {Pe'er}}, \bibinfo
  {author} {\bibfnamefont {B.}~\bibnamefont {Neyenhuis}}, \bibinfo {author}
  {\bibfnamefont {J.~J.}\ \bibnamefont {Zirbel}}, \bibinfo {author}
  {\bibfnamefont {S.}~\bibnamefont {Kotochigova}}, \bibinfo {author}
  {\bibfnamefont {P.~S.}\ \bibnamefont {Julienne}}, \bibinfo {author}
  {\bibfnamefont {D.~S.}\ \bibnamefont {Jin}}, \ and\ \bibinfo {author}
  {\bibfnamefont {J.}~\bibnamefont {Ye}},\ }\href@noop {} {\bibfield  {journal}
  {\bibinfo  {journal} {Science}\ }\textbf {\bibinfo {volume} {322}},\ \bibinfo
  {pages} {231} (\bibinfo {year} {2008})}\BibitemShut {NoStop}%
\bibitem [{\citenamefont {Lang}\ \emph {et~al.}(2008)\citenamefont {Lang},
  \citenamefont {Winkler}, \citenamefont {Strauss}, \citenamefont {Grimm},\
  and\ \citenamefont {Hecker~Denschlag}}]{Lang:ground:2008}%
  \BibitemOpen
  \bibfield  {author} {\bibinfo {author} {\bibfnamefont {F.}~\bibnamefont
  {Lang}}, \bibinfo {author} {\bibfnamefont {K.}~\bibnamefont {Winkler}},
  \bibinfo {author} {\bibfnamefont {C.}~\bibnamefont {Strauss}}, \bibinfo
  {author} {\bibfnamefont {R.}~\bibnamefont {Grimm}}, \ and\ \bibinfo {author}
  {\bibfnamefont {J.}~\bibnamefont {Hecker~Denschlag}},\ }\href@noop {}
  {\bibfield  {journal} {\bibinfo  {journal} {Phys. Rev. Lett.}\ }\textbf
  {\bibinfo {volume} {101}},\ \bibinfo {pages} {133005} (\bibinfo {year}
  {2008})}\BibitemShut {NoStop}%
\bibitem [{\citenamefont {Danzl}\ \emph {et~al.}(2010)\citenamefont {Danzl},
  \citenamefont {Mark}, \citenamefont {Haller}, \citenamefont {Gustavsson},
  \citenamefont {Hart}, \citenamefont {Aldegunde}, \citenamefont {Hutson},\
  and\ \citenamefont {N\"agerl}}]{Danzl:ground:2010}%
  \BibitemOpen
  \bibfield  {author} {\bibinfo {author} {\bibfnamefont {J.~G.}\ \bibnamefont
  {Danzl}}, \bibinfo {author} {\bibfnamefont {M.~J.}\ \bibnamefont {Mark}},
  \bibinfo {author} {\bibfnamefont {E.}~\bibnamefont {Haller}}, \bibinfo
  {author} {\bibfnamefont {M.}~\bibnamefont {Gustavsson}}, \bibinfo {author}
  {\bibfnamefont {R.}~\bibnamefont {Hart}}, \bibinfo {author} {\bibfnamefont
  {J.}~\bibnamefont {Aldegunde}}, \bibinfo {author} {\bibfnamefont {J.~M.}\
  \bibnamefont {Hutson}}, \ and\ \bibinfo {author} {\bibfnamefont {H.-C.}\
  \bibnamefont {N\"agerl}},\ }\href {\doibase doi:10.1038/nphys1533} {\bibfield
   {journal} {\bibinfo  {journal} {Nature Phys.}\ }\textbf {\bibinfo {volume}
  {6}},\ \bibinfo {pages} {265} (\bibinfo {year} {2010})}\BibitemShut {NoStop}%
\bibitem [{\citenamefont {Ospelkaus}\ \emph {et~al.}(2010)\citenamefont
  {Ospelkaus}, \citenamefont {Ni}, \citenamefont {Wang}, \citenamefont {{de
  Miranda}}, \citenamefont {Neyenhuis}, \citenamefont {Qu\'{e}m\'{e}ner},
  \citenamefont {Julienne}, \citenamefont {Bohn}, \citenamefont {Jin},\ and\
  \citenamefont {Ye}}]{Ospelkaus:react:2010}%
  \BibitemOpen
  \bibfield  {author} {\bibinfo {author} {\bibfnamefont {S.}~\bibnamefont
  {Ospelkaus}}, \bibinfo {author} {\bibfnamefont {K.-K.}\ \bibnamefont {Ni}},
  \bibinfo {author} {\bibfnamefont {D.}~\bibnamefont {Wang}}, \bibinfo {author}
  {\bibfnamefont {M.~H.~G.}\ \bibnamefont {{de Miranda}}}, \bibinfo {author}
  {\bibfnamefont {B.}~\bibnamefont {Neyenhuis}}, \bibinfo {author}
  {\bibfnamefont {G.}~\bibnamefont {Qu\'{e}m\'{e}ner}}, \bibinfo {author}
  {\bibfnamefont {P.~S.}\ \bibnamefont {Julienne}}, \bibinfo {author}
  {\bibfnamefont {J.~L.}\ \bibnamefont {Bohn}}, \bibinfo {author}
  {\bibfnamefont {D.~S.}\ \bibnamefont {Jin}}, \ and\ \bibinfo {author}
  {\bibfnamefont {J.}~\bibnamefont {Ye}},\ }\href@noop {} {\bibfield  {journal}
  {\bibinfo  {journal} {Science}\ }\textbf {\bibinfo {volume} {327}},\ \bibinfo
  {pages} {853} (\bibinfo {year} {2010})}\BibitemShut {NoStop}%
\bibitem [{\citenamefont {\ifmmode~\dot{Z}\else \.{Z}\fi{}uchowski}\ and\
  \citenamefont {Hutson}(2010)}]{Zuchowski2010}%
  \BibitemOpen
  \bibfield  {author} {\bibinfo {author} {\bibfnamefont {P.~S.}\ \bibnamefont
  {\ifmmode~\dot{Z}\else \.{Z}\fi{}uchowski}}\ and\ \bibinfo {author}
  {\bibfnamefont {J.~M.}\ \bibnamefont {Hutson}},\ }\href@noop {} {\bibfield
  {journal} {\bibinfo  {journal} {Phys. Rev. A}\ }\textbf {\bibinfo {volume}
  {81}},\ \bibinfo {pages} {060703} (\bibinfo {year} {2010})}\BibitemShut
  {NoStop}%
\bibitem [{\citenamefont {Pilch}\ \emph {et~al.}(2009)\citenamefont {Pilch},
  \citenamefont {Lange}, \citenamefont {Prantner}, \citenamefont {Kerner},
  \citenamefont {Ferlaino}, \citenamefont {N\"agerl},\ and\ \citenamefont
  {Grimm}}]{Pilch:2009}%
  \BibitemOpen
  \bibfield  {author} {\bibinfo {author} {\bibfnamefont {K.}~\bibnamefont
  {Pilch}}, \bibinfo {author} {\bibfnamefont {A.~D.}\ \bibnamefont {Lange}},
  \bibinfo {author} {\bibfnamefont {A.}~\bibnamefont {Prantner}}, \bibinfo
  {author} {\bibfnamefont {G.}~\bibnamefont {Kerner}}, \bibinfo {author}
  {\bibfnamefont {F.}~\bibnamefont {Ferlaino}}, \bibinfo {author}
  {\bibfnamefont {H.-C.}\ \bibnamefont {N\"agerl}}, \ and\ \bibinfo {author}
  {\bibfnamefont {R.}~\bibnamefont {Grimm}},\ }\href@noop {} {\bibfield
  {journal} {\bibinfo  {journal} {Phys. Rev. A}\ }\textbf {\bibinfo {volume}
  {79}},\ \bibinfo {pages} {042718} (\bibinfo {year} {2009})}\BibitemShut
  {NoStop}%
\bibitem [{\citenamefont {Debatin}\ \emph {et~al.}(2011)\citenamefont
  {Debatin}, \citenamefont {Takekoshi}, \citenamefont {Rameshan}, \citenamefont
  {Reichs\"ollner}, \citenamefont {Ferlaino}, \citenamefont {Grimm},
  \citenamefont {Vexiau}, \citenamefont {Bouloufa}, \citenamefont {Dulieu},\
  and\ \citenamefont {N\"agerl}}]{Debatin:2011}%
  \BibitemOpen
  \bibfield  {author} {\bibinfo {author} {\bibfnamefont {M.}~\bibnamefont
  {Debatin}}, \bibinfo {author} {\bibfnamefont {T.}~\bibnamefont {Takekoshi}},
  \bibinfo {author} {\bibfnamefont {R.}~\bibnamefont {Rameshan}}, \bibinfo
  {author} {\bibfnamefont {L.}~\bibnamefont {Reichs\"ollner}}, \bibinfo
  {author} {\bibfnamefont {F.}~\bibnamefont {Ferlaino}}, \bibinfo {author}
  {\bibfnamefont {R.}~\bibnamefont {Grimm}}, \bibinfo {author} {\bibfnamefont
  {R.}~\bibnamefont {Vexiau}}, \bibinfo {author} {\bibfnamefont
  {N.}~\bibnamefont {Bouloufa}}, \bibinfo {author} {\bibfnamefont
  {O.}~\bibnamefont {Dulieu}}, \ and\ \bibinfo {author} {\bibfnamefont {H.-C.}\
  \bibnamefont {N\"agerl}},\ }\href@noop {} {\bibfield  {journal} {\bibinfo
  {journal} {Phys. Chem. Chem. Phys.}\ }\textbf {\bibinfo {volume} {13}},\
  \bibinfo {pages} {18926} (\bibinfo {year} {2011})}\BibitemShut {NoStop}%
\bibitem [{\citenamefont {Takekoshi}\ \emph {et~al.}(2012)\citenamefont
  {Takekoshi}, \citenamefont {Debatin}, \citenamefont {Rameshan}, \citenamefont
  {Ferlaino}, \citenamefont {Grimm}, \citenamefont {N\"agerl}, \citenamefont
  {{Le Sueur}}, \citenamefont {Hutson}, \citenamefont {Julienne}, \citenamefont
  {Kotochigova},\ and\ \citenamefont {Tiemann}}]{Takekoshi:RbCs:2012}%
  \BibitemOpen
  \bibfield  {author} {\bibinfo {author} {\bibfnamefont {T.}~\bibnamefont
  {Takekoshi}}, \bibinfo {author} {\bibfnamefont {M.}~\bibnamefont {Debatin}},
  \bibinfo {author} {\bibfnamefont {R.}~\bibnamefont {Rameshan}}, \bibinfo
  {author} {\bibfnamefont {F.}~\bibnamefont {Ferlaino}}, \bibinfo {author}
  {\bibfnamefont {R.}~\bibnamefont {Grimm}}, \bibinfo {author} {\bibfnamefont
  {H.-C.}\ \bibnamefont {N\"agerl}}, \bibinfo {author} {\bibfnamefont {C.~R.}\
  \bibnamefont {{Le Sueur}}}, \bibinfo {author} {\bibfnamefont {J.~M.}\
  \bibnamefont {Hutson}}, \bibinfo {author} {\bibfnamefont {P.~S.}\
  \bibnamefont {Julienne}}, \bibinfo {author} {\bibfnamefont {S.}~\bibnamefont
  {Kotochigova}}, \ and\ \bibinfo {author} {\bibfnamefont {E.}~\bibnamefont
  {Tiemann}},\ }\href@noop {} {\bibfield  {journal} {\bibinfo  {journal} {Phys.
  Rev. A}\ }\textbf {\bibinfo {volume} {85}},\ \bibinfo {pages} {032506}
  (\bibinfo {year} {2012})}\BibitemShut {NoStop}%
\bibitem [{\citenamefont {McCarron}\ \emph {et~al.}(2011)\citenamefont
  {McCarron}, \citenamefont {Cho}, \citenamefont {Jenkin}, \citenamefont
  {K{\"{o}}ppinger},\ and\ \citenamefont {Cornish}}]{McCarron2011}%
  \BibitemOpen
  \bibfield  {author} {\bibinfo {author} {\bibfnamefont {D.~J.}\ \bibnamefont
  {McCarron}}, \bibinfo {author} {\bibfnamefont {H.~W.}\ \bibnamefont {Cho}},
  \bibinfo {author} {\bibfnamefont {D.~L.}\ \bibnamefont {Jenkin}}, \bibinfo
  {author} {\bibfnamefont {M.~P.}\ \bibnamefont {K{\"{o}}ppinger}}, \ and\
  \bibinfo {author} {\bibfnamefont {S.~L.}\ \bibnamefont {Cornish}},\
  }\href@noop {} {\bibfield  {journal} {\bibinfo  {journal} {Phys. Rev. A}\
  }\textbf {\bibinfo {volume} {84}},\ \bibinfo {pages} {011603} (\bibinfo
  {year} {2011})}\BibitemShut {NoStop}%
\bibitem [{\citenamefont {Cho}\ \emph {et~al.}(2011)\citenamefont {Cho},
  \citenamefont {McCarron}, \citenamefont {Jenkin}, \citenamefont
  {K{\"{o}}ppinger},\ and\ \citenamefont {Cornish}}]{Cho2011}%
  \BibitemOpen
  \bibfield  {author} {\bibinfo {author} {\bibfnamefont {H.~W.}\ \bibnamefont
  {Cho}}, \bibinfo {author} {\bibfnamefont {D.~J.}\ \bibnamefont {McCarron}},
  \bibinfo {author} {\bibfnamefont {D.~L.}\ \bibnamefont {Jenkin}}, \bibinfo
  {author} {\bibfnamefont {M.~P.}\ \bibnamefont {K{\"{o}}ppinger}}, \ and\
  \bibinfo {author} {\bibfnamefont {S.~L.}\ \bibnamefont {Cornish}},\
  }\href@noop {} {\bibfield  {journal} {\bibinfo  {journal} {Eur. Phys. J. D}\
  }\textbf {\bibinfo {volume} {65}},\ \bibinfo {pages} {125} (\bibinfo {year}
  {2011})}\BibitemShut {NoStop}%
\bibitem [{\citenamefont {Lercher}\ \emph {et~al.}(2011)\citenamefont
  {Lercher}, \citenamefont {Takekoshi}, \citenamefont {Debatin}, \citenamefont
  {Schuster}, \citenamefont {Rameshan}, \citenamefont {Ferlaino}, \citenamefont
  {Grimm},\ and\ \citenamefont {N\"agerl}}]{Lercher:2011}%
  \BibitemOpen
  \bibfield  {author} {\bibinfo {author} {\bibfnamefont {A.~D.}\ \bibnamefont
  {Lercher}}, \bibinfo {author} {\bibfnamefont {T.}~\bibnamefont {Takekoshi}},
  \bibinfo {author} {\bibfnamefont {M.}~\bibnamefont {Debatin}}, \bibinfo
  {author} {\bibfnamefont {B.}~\bibnamefont {Schuster}}, \bibinfo {author}
  {\bibfnamefont {R.}~\bibnamefont {Rameshan}}, \bibinfo {author}
  {\bibfnamefont {F.}~\bibnamefont {Ferlaino}}, \bibinfo {author}
  {\bibfnamefont {R.}~\bibnamefont {Grimm}}, \ and\ \bibinfo {author}
  {\bibfnamefont {H.-C.}\ \bibnamefont {N\"agerl}},\ }\href {\doibase
  10.1140/epjd/e2011-20015-6} {\bibfield  {journal} {\bibinfo  {journal} {Eur.
  Phys. J. D}\ }\textbf {\bibinfo {volume} {65}},\ \bibinfo {pages} {3}
  (\bibinfo {year} {2011})}\BibitemShut {NoStop}%
\bibitem [{\citenamefont {Lin}\ \emph {et~al.}(2009)\citenamefont {Lin},
  \citenamefont {Perry}, \citenamefont {Compton}, \citenamefont {Spielman},\
  and\ \citenamefont {Porto}}]{Lin2009}%
  \BibitemOpen
  \bibfield  {author} {\bibinfo {author} {\bibfnamefont {Y.-J.}\ \bibnamefont
  {Lin}}, \bibinfo {author} {\bibfnamefont {A.~R.}\ \bibnamefont {Perry}},
  \bibinfo {author} {\bibfnamefont {R.~L.}\ \bibnamefont {Compton}}, \bibinfo
  {author} {\bibfnamefont {I.~B.}\ \bibnamefont {Spielman}}, \ and\ \bibinfo
  {author} {\bibfnamefont {J.~V.}\ \bibnamefont {Porto}},\ }\href@noop {}
  {\bibfield  {journal} {\bibinfo  {journal} {Phys. Rev. A}\ }\textbf {\bibinfo
  {volume} {79}},\ \bibinfo {pages} {063631} (\bibinfo {year}
  {2009})}\BibitemShut {NoStop}%
\bibitem [{\citenamefont {Jenkin}\ \emph {et~al.}(2011)\citenamefont {Jenkin},
  \citenamefont {McCarron}, \citenamefont {K{\"{o}}ppinger}, \citenamefont
  {Cho}, \citenamefont {Hopkins},\ and\ \citenamefont {Cornish}}]{Jenkin2011}%
  \BibitemOpen
  \bibfield  {author} {\bibinfo {author} {\bibfnamefont {D.~L.}\ \bibnamefont
  {Jenkin}}, \bibinfo {author} {\bibfnamefont {D.~J.}\ \bibnamefont
  {McCarron}}, \bibinfo {author} {\bibfnamefont {M.~P.}\ \bibnamefont
  {K{\"{o}}ppinger}}, \bibinfo {author} {\bibfnamefont {H.~W.}\ \bibnamefont
  {Cho}}, \bibinfo {author} {\bibfnamefont {S.~A.}\ \bibnamefont {Hopkins}}, \
  and\ \bibinfo {author} {\bibfnamefont {S.~L.}\ \bibnamefont {Cornish}},\
  }\href@noop {} {\bibfield  {journal} {\bibinfo  {journal} {Eur. Phys. J. D}\
  }\textbf {\bibinfo {volume} {65}},\ \bibinfo {pages} {11} (\bibinfo {year}
  {2011})}\BibitemShut {NoStop}%
\bibitem [{Note1()}]{Note1}%
  \BibitemOpen
  \bibinfo {note} {Units of gauss rather than tesla, the accepted SI unit for
  the magnetic field, have been used in this paper to conform to the
  conventional usage in this field of physics.}\BibitemShut {Stop}%
\bibitem [{\citenamefont {Wille}\ \emph {et~al.}(2008)\citenamefont {Wille},
  \citenamefont {Spiegelhalder}, \citenamefont {Kerner}, \citenamefont {Naik},
  \citenamefont {Trenkwalder}, \citenamefont {Hendl}, \citenamefont {Schreck},
  \citenamefont {Grimm}, \citenamefont {Tiecke}, \citenamefont {Walraven},
  \citenamefont {Kokkelmans}, \citenamefont {Tiesinga},\ and\ \citenamefont
  {Julienne}}]{Wille2008}%
  \BibitemOpen
  \bibfield  {author} {\bibinfo {author} {\bibfnamefont {E.}~\bibnamefont
  {Wille}}, \bibinfo {author} {\bibfnamefont {F.~M.}\ \bibnamefont
  {Spiegelhalder}}, \bibinfo {author} {\bibfnamefont {G.}~\bibnamefont
  {Kerner}}, \bibinfo {author} {\bibfnamefont {D.}~\bibnamefont {Naik}},
  \bibinfo {author} {\bibfnamefont {A.}~\bibnamefont {Trenkwalder}}, \bibinfo
  {author} {\bibfnamefont {G.}~\bibnamefont {Hendl}}, \bibinfo {author}
  {\bibfnamefont {F.}~\bibnamefont {Schreck}}, \bibinfo {author} {\bibfnamefont
  {R.}~\bibnamefont {Grimm}}, \bibinfo {author} {\bibfnamefont {T.~G.}\
  \bibnamefont {Tiecke}}, \bibinfo {author} {\bibfnamefont {J.~T.~M.}\
  \bibnamefont {Walraven}}, \bibinfo {author} {\bibfnamefont {S.~J. J. M.~F.}\
  \bibnamefont {Kokkelmans}}, \bibinfo {author} {\bibfnamefont
  {E.}~\bibnamefont {Tiesinga}}, \ and\ \bibinfo {author} {\bibfnamefont
  {P.~S.}\ \bibnamefont {Julienne}},\ }\href@noop {} {\bibfield  {journal}
  {\bibinfo  {journal} {Phys. Rev. Lett.}\ }\textbf {\bibinfo {volume} {100}},\
  \bibinfo {pages} {053201} (\bibinfo {year} {2008})}\BibitemShut {NoStop}%
\bibitem [{\citenamefont {Blackley}\ \emph {et~al.}()\citenamefont {Blackley}
  \emph {et~al.}}]{Blackley2012}%
  \BibitemOpen
  \bibfield  {author} {\bibinfo {author} {\bibfnamefont {C.~L.}\ \bibnamefont
  {Blackley}} \emph {et~al.},\ }\href@noop {} {\bibinfo  {journal} {in
  preparation}\ }\BibitemShut {NoStop}%
\bibitem [{\citenamefont {Hutson}\ and\ \citenamefont
  {Green}(1994)}]{molscat:v14-short}%
  \BibitemOpen
\bibfield  {journal} {  }\bibfield  {author} {\bibinfo {author} {\bibfnamefont
  {J.~M.}\ \bibnamefont {Hutson}}\ and\ \bibinfo {author} {\bibfnamefont
  {S.}~\bibnamefont {Green}},\ }\href@noop {} {\emph {\bibinfo {title}
  {{MOLSCAT} computer program, version 14}}}\ (\bibinfo  {publisher} {CCP6},\
  \bibinfo {address} {Daresbury},\ \bibinfo {year} {1994})\BibitemShut
  {NoStop}%
\bibitem [{\citenamefont {Gonz\'{a}lez-Mart\'{\i}nez}\ and\ \citenamefont
  {Hutson}(2007)}]{Gonzalez-Martinez:2007}%
  \BibitemOpen
  \bibfield  {author} {\bibinfo {author} {\bibfnamefont {M.~L.}\ \bibnamefont
  {Gonz\'{a}lez-Mart\'{\i}nez}}\ and\ \bibinfo {author} {\bibfnamefont {J.~M.}\
  \bibnamefont {Hutson}},\ }\href@noop {} {\bibfield  {journal} {\bibinfo
  {journal} {Phys. Rev. A}\ }\textbf {\bibinfo {volume} {75}},\ \bibinfo
  {pages} {022702} (\bibinfo {year} {2007})}\BibitemShut {NoStop}%
\bibitem [{\citenamefont {Manolopoulos}(1986)}]{Manolopoulos:1986}%
  \BibitemOpen
  \bibfield  {author} {\bibinfo {author} {\bibfnamefont {D.~E.}\ \bibnamefont
  {Manolopoulos}},\ }\href@noop {} {\bibfield  {journal} {\bibinfo  {journal}
  {J. Chem. Phys.}\ }\textbf {\bibinfo {volume} {85}},\ \bibinfo {pages} {6425}
  (\bibinfo {year} {1986})}\BibitemShut {NoStop}%
\bibitem [{\citenamefont {Alexander}(1984)}]{Alexander:1984}%
  \BibitemOpen
  \bibfield  {author} {\bibinfo {author} {\bibfnamefont {M.~H.}\ \bibnamefont
  {Alexander}},\ }\href@noop {} {\bibfield  {journal} {\bibinfo  {journal} {J.
  Chem. Phys.}\ }\textbf {\bibinfo {volume} {81}},\ \bibinfo {pages} {4510}
  (\bibinfo {year} {1984})}\BibitemShut {NoStop}%
\bibitem [{\citenamefont {Hutson}(2007)}]{Hutson:res:2007}%
  \BibitemOpen
  \bibfield  {author} {\bibinfo {author} {\bibfnamefont {J.~M.}\ \bibnamefont
  {Hutson}},\ }\href@noop {} {\bibfield  {journal} {\bibinfo  {journal} {New J.
  Phys.}\ }\textbf {\bibinfo {volume} {9}},\ \bibinfo {pages} {152} (\bibinfo
  {year} {2007})}\BibitemShut {NoStop}%
\bibitem [{\citenamefont {Hutson}(1993)}]{Hutson:bound-short:1993}%
  \BibitemOpen
  \bibfield  {author} {\bibinfo {author} {\bibfnamefont {J.~M.}\ \bibnamefont
  {Hutson}},\ }\href@noop {} {\emph {\bibinfo {title} {{BOUND} computer
  program, version 5}}}\ (\bibinfo  {publisher} {CCP6},\ \bibinfo {address}
  {Daresbury},\ \bibinfo {year} {1993})\BibitemShut {NoStop}%
\bibitem [{\citenamefont {Hutson}\ \emph {et~al.}(2008)\citenamefont {Hutson},
  \citenamefont {Tiesinga},\ and\ \citenamefont {Julienne}}]{Hutson:Cs2:2008}%
  \BibitemOpen
  \bibfield  {author} {\bibinfo {author} {\bibfnamefont {J.~M.}\ \bibnamefont
  {Hutson}}, \bibinfo {author} {\bibfnamefont {E.}~\bibnamefont {Tiesinga}}, \
  and\ \bibinfo {author} {\bibfnamefont {P.~S.}\ \bibnamefont {Julienne}},\
  }\href@noop {} {\bibfield  {journal} {\bibinfo  {journal} {Phys. Rev. A}\
  }\textbf {\bibinfo {volume} {78}},\ \bibinfo {pages} {052703} (\bibinfo
  {year} {2008})}\BibitemShut {NoStop}%
\bibitem [{Note2()}]{Note2}%
  \BibitemOpen
  \bibinfo {note} {Comparison between theory and experiment for the cluster of
  resonances near 110~G is limited by the fact that not all resonances are
  resolved experimentally. Two experimental observations, strong thermalisation
  at 233.9(2)~G and strong loss at 246.5(3)~G, do not correspond to 2-body
  resonances using the current theoretical model and require further
  investigation.}\BibitemShut {Stop}%
\bibitem [{Note3()}]{Note3}%
  \BibitemOpen
  \bibinfo {note} {The two $p$-wave resonances that are observed experimentally
  have calculated widths $>6$~G, and all others have calculated widths
  $<0.3$~G.}\BibitemShut {Stop}%
\end{thebibliography}%

\end{document}